\documentclass[fleqn]{revtex4}

\usepackage{amsmath}
\usepackage{verbatim}

\begin{document}

\title{Nonrelativistic energy levels of helium atom}

\author{D.T.~Aznabaev$^{1,2,3}$}
\author{A.K.~Bekbaev$^{1,4}$}
\author{Vladimir I. Korobov$^{1,5}$}
\affiliation{$^1$Bogoliubov Laboratory of Theoretical Physics, Joint Institute
for Nuclear Research, Dubna 141980, Russia,}
\affiliation{$^2$The Institute of Nuclear Physics, Ministry of Energy of the Republic of Kazakhstan, 050032 Almaty, Kazakhstan}
\affiliation{$^3$L.N.~Gumilyov Eurasian National University, 010000 Astana, Kazakhstan}
\affiliation{$^4$Al-Farabi Kazakh National University, 050038 Almaty, Kazakhstan}
\affiliation{$^5$Peoples' Friendship University of Russia (RUDN University), 6 Miklukho-Maklaya St, Moscow, 117198, Russia}


\begin{abstract}
The nonrelativistic ionization energy levels of a helium atom are calculated for $S$, $P$, $D$ and $F$ states. The calculations are based on the variational method of "exponential" expansion. The convergence of the calculated energy levels is studied as a function of the number of basis functions $N$. This allows us to claim that the obtained energy values (including the values for the states with a nonzero angular momentum) are accurate up to 28-35 significant digits. Calculations of the nonrelativistic ionization energy of the negative hydrogen ion H$^-$ are also presented.
\end{abstract}

\maketitle

\section{Introduction}

The quantum problem of three bodies with Coulomb interaction is one of the most notable nonintegrable problems in  quantum mechanics.  At  the  same time, extremely accurate numerical solutions for the problem of bound states for a system of three particles may be obtained with modern computers. For example, the nonrelativistic energy of the ground state of helium with a nucleus of an infinite mass is now known accurately to 46 significant digits \cite{Sch06}.

In the present study, a version of the variational method (the so called "exponential" expansion) that allows to numerically solve the quantum Coulomb three-body bound state problem with a very high precision, which is easily applicable as well to the states with a nonzero angular momentum, is considered. This method is used to calculate the nonrelativistic ionization energies of a helium atom for $S$, $P$, $D$, and $F$ states. It is shown that the developed method is an efficient and flexible instrument for studying Coulomb systems. An analysis of convergence proves that the method is highly accurate and demonstrates that nonrelativistic energies accurate up to 28-35 significant digits may be obtained with rather moderate efforts.

Developing of such high precision methods is of importance for the reason that it may help solving a wide variety of problems that are of interest in physics. For example, antiprotonic helium atoms are of particular interest, which allows for high precision studies of energy spectrum of this exotic system and inferring of various properties of an antiproton from comparison of theory and experiment \cite{Nature,Korobov14}. Here it is worthy to mention a recent interest to the antiprotonic helium as a tool for constrains on various fifth forces \cite{Salumbides,Ficek} to set general limits on new interactions beyond the Standard Model.

Another important aspect, namely, the cross impact of atomic and nuclear physics \cite{Friar}, in the determination of statistical parameters of nuclei should be noted. For example, the accuracy of the mean square helium charge radius that is determined experimentally  from  electron-nuclei scattering is about 1--3\%. At the same time, the experimental determination of the charge radius of $^4$He by muonic atom spectroscopy \cite{Pohl11} allows one to reduce the error in the value of this parameter by more than an order of magnitude.

One more application is the numerical-analytical studies of the critical nuclear charge $Z_c$ for two-electron atoms \cite{Drake14} and the $1/Z$ expansion of the binding energy \cite{Morgan}, where the high precision calculations are extremely crucial.

The paper is structured as follows. The variational principle and application of the variational Ritz method to the stationary Schr\"odinger problem is discussed in detail in Sections 1 and 2. Particularly, the variational "exponential"  expansion used in practical calculations is formulated. The inverse  iteration  method,  which  is considered to be one of the most efficient computational tools to solve an eigenvalue problem for a finite basis, is presented in Section 3. In the last section, the numerical calculations are reviewed, and the final theoretical results for 19 states of a helium atom as well as the most accurate so far estimate for the ground state energy of H$^-$ ion are given.


\section{Variational method}

Let us first formulate the variational principle for bound states.

The Hylleraas-Undheim variational principle, which is better known in mathematics as the Rayleigh-Ritz variational principle, is the starting point in solving the stationary Schrodinger equation,
\begin{equation}
H\psi=E\psi
\end{equation}
for a certain Hamiltonian using variational methods. This principle is considered a versatile method for deriving an  approximate solution. The problems of determining the extrema or stationary values of functionals are the basic problems of variational calculation. The essence of this method consists of substituting the problem of finding the stationary  values of functionals with a fundamentally less complex problem of finding the stationary values of functions of several variables \cite{ReedSimon}.

Let there be a self-adjoint operator defined within the Hilbert space for which the following boundedness
condition is satisfied:
\begin{equation}\label{cond}
    H\geq{cI},
\end{equation}
where $c$ is a certain constant. Let us then define a functional
\begin{equation}
\Phi(\psi)=\frac{(\psi,H\psi)}{(\psi,\psi)},
\end{equation}
that is bounded from below by $c$. The following theorem is valid:

\vspace{3mm}

\textbf{Theorem} \cite{ReedSimon}. Let $H$ be a self-adjoint operator that satisfies condition (\ref{cond}). Let us define
\newcommand{\chim}{\raisebox{1pt}{$\scriptstyle\chi$}}
\begin{equation}\label{minmax}
\mu_n(H)=\max_{\dim \chim=n-1}\;
         \min_{\genfrac{}{}{0pt}{}{\Psi\in\mathcal{D}(H)}{\Psi\in\chim^\bot}}
\Phi(\Psi),
\end{equation}
where $\chi^\bot$ is a subspace orthogonal to  $\chi$, $D(H)$ is the domain of operator $H$. One of the following assertions is then true for any fixed $n$:

$a)$ $n$ eigenvalue (degenerate eigenvalues are counted according to their multiplicity) lying below the essential  spectrum  boundary  are  present and $\mu_n (H)$ is the $n$-th eigenvalue (with account of multiplicity); or

$b)$ $\mu_n (H)$ is the lower boundary of the essential spectrum.

\vspace{3mm}

The determination of eigenvalues (i.e., the energy of bound states of the stationary Schrodinger equation) comes down to  calculating the saddle points of functional (\ref{minmax}). The assertion of the theorem is known as the minimax principle.

Let us now consider a method that uses the Rayleigh--Ritz variational principle to solve practical eigenvalue problems and is called the Ritz process. Let $\phi_k$ be a complete sequence of vectors in the Hilbert space subject to the following conditions:
\begin{enumerate}\itemsep 1mm
\item  vectors $\phi_k$ belong to the domain of operator  $H$;

\item  vectors $\phi_1,\phi_2,...,\phi_n$ are linearly independent at any $n$.
\end{enumerate}

Let us assume that $u_n=\sum_{k=1}^n x_k\phi_k$, where $x_k$ are scalar coefficients. Inserting $u_n$  (at fixed $n$ into functional $\Phi(\cdot)$, one obtains a function that depends on a finite set of parameters  $\{x_n\}_1^n$:
\begin{equation}
    \Phi(x)=\biggl(~\sum\limits_{i,j=1}^n a_{ij}x_ix_j\biggl)\biggl{/}\biggl(~\sum\limits_{i,j=1}^n b_{ij}x_ix_j\biggl),
\end{equation}
where
\begin{equation}
    a_{ij}=(\phi_i,H\phi_j), \qquad
    b_{ij}=(\phi_i,\phi_j).
\end{equation}

The determination of minimax solutions is thus reduced to calculating the corresponding eigenvalues of the generalized eigenvalue problem:
\begin{equation}\label{gep}
    Ax=\lambda Bx
\end{equation}
where matrices $A$ and $B$ are composed of coefficients $a_{ij}$ and $b_{ij}$, respectively.

Vectors $\phi_k$ may depend on nonlinear parameters $\omega$. If this is the case, problem (\ref{gep}) is solved for each fixed $\omega$ and each eigenvalue number $k$, $\lambda_k (\omega)$ is chosen, and this value is then minimized over all values of nonlinear parameters:
\begin{equation}
\lambda_k=\inf_{\omega} \lambda_k (\omega).
\end{equation}

One  important  condition  is  satisfied  for  Ritz  estimates:
\begin{equation}\label{varb}
   \mu_k (H)\leq \lambda_k.
\end{equation}
It follows from there that Ritz estimates are upper bounds. Inequality (\ref{varb}) for basis functions dependent on nonlinear parameters follows from:
\begin{equation}
   \mu_k (H)\leq \inf_{\omega} \lambda_k (\omega)=\lambda_k.
\end{equation}
A rigorous proof of the applicability of Theorem 1 to the problems of nonrelativistic quantum mechanics with a Hamiltonian of the form:
\begin{equation}
H = -\sum\limits_{i=1}^n \frac{\Delta_i}{2m_i}+V(\mathbf{r}_1,...,\mathbf{r}_n),
\end{equation}
and a potential of a sufficiently general form that includes, among others, the Coulomb potential of interparticle interaction was derived by Kato \cite{Kato}.

\section{Generalized Hylleraas expansion.}

Let us consider the generalized Hylleraas expansion for the states of arbitrary total orbital momentum $L$ \cite{Drake}:
\begin{equation}\label{GenHyl}
\psi(r_1,r_2)=\sum_{l_1+l_2=\mathcal{L}}\mathcal{Y}_{LM}^{l_1 l_2}(r_1,r_2)\biggl[e^{-\alpha r_1-\beta r_2-\gamma r_{12}}\sum_{l,m,n\geq0}C_{lmn}
r_1^l r_2^m r_{12}^n\biggl],
\end{equation}
where $\mathcal{L}=L$ for the states of "normal" spatial parity $\Pi=(-1)^L$, and $\mathcal{L}=L+1$ for the states of "anomalous" spatial parity $\Pi=(-1)^{L+1}$. The $\mathcal{Y}_{LM}^{l_1 l_2}$ functions are regular bipolar spherical harmonics \cite{Varsh} that depend on two angular coordinates:
\[
\mathcal{Y}_{LM}^{l_1 l_2}(r_1 r_2)=r_1^{l_1} r_2^{l_2}\{Y_{l_1}(\hat{r}_1)\otimes Y_{l_2}(\hat{r}_2)\}_{LM},
\]
and spatial parity operator $P\psi=\pi\psi$,  acts on the spatial coordinates in the following way:  $P(r_1,r_2)\rightarrow (-r_1,-r_2)$. The ease of use of the $\mathcal{Y}_{LM}^{l_1 l_2}$  functions stems from the fact that they correctly reproduce the behavior of the wave function at  $r_1\rightarrow 0$  (or $r_2\rightarrow 0$ ), and retain the reasonable  requirement  of  boundedness  of  the  function within the domain of variables for the expression within square brackets in Eq. (\ref{GenHyl}).

The "normal" and "anomalous" spatial parities were designated this way for the following reason. It can be seen from expansion (\ref{GenHyl}) that "anomalous" parity states in a dissociation limit may be decomposed into clusters with a bound pair, which may have only nonzero angular momentum, $l\geq 1$. In atomic physics, the ground state of a pair of particles has zero angular momentum, while the boundary of the continuum in a system of three particles is defined by the energy of the pair with the lowest ground energy, or zero (if no bound pairs are presented). It follows that bound "anomalous" parity states are normally located in the continuum of the three particle system \cite{HZhW}. Therefore, these states are imbedded into the continuum and any perturbation violating spatial parity of the system makes them "true" resonances.

The calculation of matrix elements reduces to evaluating integrals of the following form:
\begin{equation}\label{integral}
\Gamma_{lmn}(\alpha,\beta,\gamma) =
   \int\int r_1^l r_2^m r_{12}^n e^{-\alpha r_1-\beta r_2-\gamma r_{12}}dr_1 dr_2 dr_{12}.
\end{equation}
Differentiating  with respect to $\alpha$ under the integral sign, we obtain the following:
\begin{equation}
\nonumber
\biggl(-\frac{\partial}{\partial\alpha}\biggl)\Gamma_{l-1,mn}(\alpha,\beta,\gamma)=\Gamma_{lmn}(\alpha,\beta,\gamma),
\end{equation}
Thus, all integrals may be evaluated from $\Gamma_{000}$ by simple differentiation:
\begin{equation}\label{master}
\begin{array}{@{}l}\displaystyle
    \Gamma_{lmn}(\alpha,\beta,\gamma)=\biggl(-\frac {\partial}{\partial\alpha}\biggl)^l \biggl(-\frac {\partial}{\partial\beta}\biggl)^m \biggl(-\frac {\partial}{\partial\gamma}\biggl)^n \times\Gamma_{000}(\alpha,\beta,\gamma)
\\[3mm]\displaystyle\hspace{20mm}
    =\biggl(-\frac {\partial}{\partial\alpha}\biggl)^l \biggl(-\frac {\partial}{\partial\beta}\biggl)^m \biggl(-\frac {\partial}{\partial\gamma}\biggl)^n
    \biggl[\frac{2}{(\alpha+\beta)(\beta+\gamma)(\gamma+\alpha)}\biggl].
\end{array}
\end{equation}

Following \cite{Sack}, we then use recurrence relation
\begin{equation}
\nonumber
    \Gamma_{lm}(\alpha,\beta)=\frac {1} {\alpha+\beta}\biggl[\biggl(l\Gamma_{l-1,m}+m\Gamma_{l,m-1}\biggl)+\biggl(-\frac{\partial}{\partial\alpha}\biggl)^l\biggl(-\frac{\partial}{\partial\beta}\biggl)^m f(\alpha,\beta)\biggl].
\end{equation}
Applying it successively to each pair of variables $\alpha$, $\beta$ and $\gamma$,  we  arrive  at  the  recurrence  scheme  for  integral evaluation for nonnegative values of parameters ($(l,m,n)$:
\begin{equation}\label{recu}
\begin{array}{@{}l}\displaystyle
    \Gamma_{lmn}=\frac {1}{\alpha+\beta}\biggl[l\Gamma_{l-1,m,n}+m\Gamma_{l,m-1,n}+B_{lmn}\biggl],
\\[5mm]\displaystyle
    B_{lmn}=\frac {1}{\alpha+\beta}\biggl[lB_{l-1,m,n}+mB_{l,m-1,n}+A_{lmn}\biggl],
\\[5mm]\displaystyle
    A_{lmn}=\delta_{l0}\frac {2(m+n)!}{(\beta+\gamma)^{m+n+1}}.
\end{array}
\end{equation}

The fact that quantities $A_{lmn}$, $B_{lmn}$, and $\Gamma_{lmn}$ in (\ref{recu}) are positive is an important feature of these relations  that  makes  recurrence scheme (\ref{recu}) being stable with respect to the rounding off errors in computations. Averaging over angular variables for the states with a nonzero total orbital moment of the system was analyzed by Drake \cite{Drake}. This averaging reduces the calculation of matrix elements for nonzero $L$ to integrals of the type (\ref{integral}). A compact and efficient recurrence scheme that implements this reduction was proposed later by Efros \cite{Efros}. The  efficiency  of  the  above described  variational expansions is the highest when they are applied to systems composed of two electrons and a heavy nucleus.

Let us now study "exponential" expansion in more details. This expansion assumes the following form for $S$ states:
\begin{equation}\label{varexp}
    \psi(r_1,r_2,r_{12})=\sum_n C_ne^{-\alpha_n r_1-\beta_n r_2-\gamma_n r_{12}},
\end{equation}
where  the  parameters  in  the  exponent  are  chosen  in one  way  or  another.  In early studies \cite{Ros71} that  used expansion  (\ref{varexp}), the obtained  representation  was associated with the discretization of the integral representation of the wave function
\begin{equation}\label{quadrature}
    \psi(x_1,...,x_A)=\int\varphi(x_1,...,x_A;\alpha)f(\alpha)d\alpha,
\end{equation}
that was proposed by Griffin and Wheeler \cite{Wheeler} in 1957. The $\alpha_n$  $\beta_n$ and $\gamma_n$ parameters were chosen in accordance with various quadrature integration formulas (\ref{quadrature}). The systematic study of expansion (\ref{varexp}) with parameters generated using  pseudorandom numbers was carried out in \cite{Smith}. In the proposed approach, nonlinear parameters from Eq.~(\ref{varexp}) are generated using the following simple formulas:
\begin{equation}\label{gener}
\begin{array}{@{}l}\displaystyle
    \alpha_n=\biggl[\biggl\lfloor\frac{1}{2}n(n+1)\sqrt{p_\alpha}\biggl\rfloor(A_2-A_1)+A_1\biggl],
\\[4mm]\displaystyle
    \beta_n =\biggl[\biggl\lfloor\frac{1}{2}n(n+1)\sqrt{p_\beta} \biggl\rfloor(B_2-B_1)+B_1\biggl],
\\[4mm]\displaystyle
    \gamma_n=\biggl[\biggl\lfloor\frac {1}{2}n(n+1)\sqrt{p_\gamma}\biggl\rfloor(C_2-C_1)+C_1\biggl],
\end{array}
\end{equation}
where $\lfloor x\rfloor$ is the fractional part of $x$ and $p_\alpha, p_\beta$, and $p_\gamma$ are certain prime numbers. These simple generators of pseudorandom numbers have their advantage in the reproducibility of the results of  variational  calculations. The convergence rate of the exponential expansion with a pseudorandom strategy for choosing nonlinear parameters (\ref{gener}) is exceptionally high at the sets of basis functions of moderate dimensionalities (up to 100--200 test functions). Rapid basis degeneration that results in the loss of computational stability in the double precision arithmetic by basis dimensionality $N$=200 is among the disadvantages of the method.

\begin{table}[t]
\caption{Convergence of the nonrelativistic energy of the ground state of a helium atom.} \label{converg1}
\begin{center}
\begin{tabular}{c@{\hspace{10mm}}l@{\hspace{10mm}}r}
\hline\hline
\vrule width0pt height 11pt
Basis $(N)$ & $E_{nr}$ \\
\hline
\vrule width0pt height 12pt
 10000  & $-$2.90372\,43770\,34119\,59831\,11592\,45193\,9 \\
 14000  & $-$2.90372\,43770\,34119\,59831\,11592\,45194\,398 \\
 18000  & $-$2.90372\,43770\,34119\,59831\,11592\,45194\,40432  \\
 22000  & $-$2.90372\,43770\,34119\,59831\,11592\,45194\,40443  \\
\hline\hline
\end{tabular}
\end{center}
\end{table}

Let us write out for convenience  the  exponential variational expansion in its complete form, which accounts for the angular dependence of the wave function that describes the rotational degrees of freedom:
\begin{equation}\label{exp_main}
\begin{array}{@{}l}\displaystyle
\Psi(\mathbf{r}_1,\mathbf{r}_2) =
       \sum_{l_1+l_2=\mathcal{L}}
         \mathcal{Y}^{l_1l_2}_{LM}(\hat{\mathbf{r}}_1,\hat{\mathbf{r}}_2)
         G^{L\pi}_{l_1l_2}(r_1,r_2,r_{12}),
\\[6mm]\displaystyle
 G^{L\pi}_{l_1l_2}(r_1,r_2,r_{12}) =
     \sum_n C_n\> e^{-\alpha_n r_1-\beta_n r_2-\gamma_n r_{12}},
\end{array}
\end{equation}
where $\mathcal{L}=L$ or $L\!+\!1$ (depending on the spatial parity of the state) and the complex parameters in the exponent are generated in a pseudorandom way (\ref{gener}).

When the number of basis function $N$ increases, one may observe that the convergence rate is slowing down. This may be attributed to the fact that the "exact" wave function of the bound state for atomic helium has a logarithmic singularity at $r_1,r_2\to0$: $\rho^2\ln{\rho}$, where $\rho=\sqrt{r_1^2+r_2^2}$ is the hyperradius of two electrons \cite{Fock}.  In order to improve the situation, one should construct a multilayer variational expansion composed  of  several  independent  sets  of  basis  functions, the optimal variational nonlinear parameters for which are found independently. Thus, each set of basis functions defines the best approximation in a certain region of coordinates of the system. In the case of a helium atom, the regions should be enclosed within each other and be more and more compact in terms of the hyperradius ($\rho<\rho_n=a^n$, where $a\approx0.1$ and $n=1,2,3\dots$). This strategy makes the exponential expansion an efficient and versatile computational method for bound states in the quantum three-body problem with Coulomb interaction. The capabilities of this method were demonstrated in \cite{Korobov00,Korobov02}.

\begin{table}[t]
\caption{Nonrelativistic energies of the $S$, $P$, $D$, and $F$ states of a helium atom. $N$ is the number of basis functions. The two lines represent two consecutive calculations with the largest basis sets to show convergent digits. The third line presents calculations by Drake and Yan \cite{Drake92}.}\label{energies}
\begin{center}
\begin{tabular}{c@{\hspace{2.5mm}}c@{\hspace{2.5mm}}l@{\hspace{5mm}}c@{\hspace{2.5mm}}c@{\hspace{2.5mm}}l}
\hline\hline
\vrule width0pt height 11pt
State     & $N$ & ~~~~~~~~~$E_{nr}$ & State     & $N$ & ~~~~~~~~~$E_{nr}$   \\
\hline
\vrule width0pt height 12pt
 $1^1S$ & 18000  & $-2.90372\>43770\>34119\>59831\>11592\>45194\>40432$  & $4^1S$ & 14000  & $-2.03358\>67170\>30725\>44743\>92926\>44363\>64$ \\
 $1^1S$ & 22000  & $-2.90372\>43770\>34119\>59831\>11592\>45194\>40443$  & $4^1S$ & 18000  & $-2.03358\>67170\>30725\>44743\>92926\>44363\>87$ \\[2mm]
 $2^1S$ & 18000  & $-2.14597\>40460\>54417\>41580\>50289\>75461\>918$    & $4^3S$ & 14000  & $-2.03651\>20830\>98236\>29958\>03780\>71617\>853$ \\
 $2^1S$ & 22000  & $-2.14597\>40460\>54417\>41580\>50289\>75461\>921$    & $4^3S$ & 16000  & $-2.03651\>20830\>98236\>29958\>03780\>71617\>874$ \\
 & \cite{Drake92}& $-2.14597\>40460\>5443(5)$  & \\[1mm]
 $2^3S$ & 14000  & $-2.17522\>93782\>36791\>30573\>89782\>78206\>81124$  & $4^1P$ & 18000  & $-2.03106\>96504\>50240\>71475\>89314\>36090\>3$ \\
 $2^3S$ & 16000  & $-2.17522\>93782\>36791\>30573\>89782\>78206\>81125$  & $4^1P$ & 22000  & $-2.03106\>96504\>50240\>71475\>89314\>36094\>1$ \\
 & \cite{Drake92}& $-2.17522\>93782\>367912(1)$  & & \cite{Drake92} & $-2.03106\>96504\>5024(3)$ \\[1mm]
 $2^1P$ & 18000  & $-2.12384\>30864\>98101\>35924\>73331\>42354$         & $4^3P$ & 18000  & $-2.03232\>43542\>96630\>33195\>38824\>67087$  \\
 $2^1P$ & 22000  & $-2.12384\>30864\>98101\>35924\>73331\>42374$         & $4^3P$ & 22000  & $-2.03232\>43542\>96630\>33195\>38824\>67103$  \\
 & \cite{Drake92}& $-2.12384\>30864\>98092(8)$  & & \cite{Drake92} & $-2.03232\>43542\>9662(2)$ \\[1mm]
 $2^3P$ & 16000  & $-2.13316\>41907\>79283\>20514\>69927\>63793$         & $4^1D$ & 22000  & $-2.03127\>98461\>78684\>99621\>39438\>073$     \\
 $2^3P$ & 18000  & $-2.13316\>41907\>79283\>20514\>69927\>63806$         & $4^1D$ & 26000  & $-2.03127\>98461\>78684\>99621\>39438\>143$     \\
 & \cite{Drake92}& $-2.13316\>41907\>7927(1)$  & & \cite{Drake92} & $-2.03127\>98461\>78687(7)$ \\[1mm]
 $3^1S$ & 18000  & $-2.06127\>19897\>40908\>65074\>03499\>37089\>2816$   & $4^3D$ & 18000  & $-2.03128\>88475\>01795\>53802\>34920\>591$     \\
 $3^1S$ & 22000  & $-2.06127\>19897\>40908\>65074\>03499\>37089\>2824$   & $4^3D$ & 22000  & $-2.03128\>88475\>01795\>53802\>34920\>630$     \\[1mm]
 & & & & \cite{Drake92}& $-2.03128\>88475\>01795(3)$  \\[1mm]
 $3^3S$ & 14000  & $-2.06868\>90674\>72457\>19199\>65329\>11291\>75048$  & $4^1F$ & 18000  & $-2.03125\>51443\>81748\>60863\>20824\>071$    \\
 $3^3S$ & 16000  & $-2.06868\>90674\>72457\>19199\>65329\>11291\>75049$  & $4^1F$ & 22000  & $-2.03125\>51443\>81748\>60863\>20824\>079$    \\[1mm]
 & & & & \cite{Drake92}& $-2.03125\>51443\>81749(1)$  \\[1mm]
 $3^1P$ & 18000  & $-2.05514\>63620\>91943\>53692\>83410\>913$           & $4^3F$ & 18000  & $-2.03125\>51684\>03245\>39350\>49887\>2817$   \\
 $3^1P$ & 22000  & $-2.05514\>63620\>91943\>53692\>83410\>921$           & $4^3F$ & 22000  & $-2.03125\>51684\>03245\>39350\>49887\>2846$   \\
 & \cite{Drake92}& $-2.05514\>63620\>9195(3)$  & & \cite{Drake92}& $-2.03125\>51684\>032454(6)$  \\[1mm]
 $3^3P$ & 18000  & $-2.05808\>10842\>74275\>33134\>26965\>47197$         \\
 $3^3P$ & 22000  & $-2.05808\>10842\>74275\>33134\>26965\>47203$         \\
 & \cite{Drake92}& $-2.05808\>10842\>7428(4)$  & \\[1mm]
 $3^1D$ & 18000  & $-2.05562\>07328\>52246\>48939\>00994\>819$           \\
 $3^1D$ & 22000  & $-2.05562\>07328\>52246\>48939\>00994\>825$           \\
 & \cite{Drake92}& $-2.05562\>07328\>52245(6)$  & \\[1mm]
 $3^3D$ & 18000  & $-2.05563\>63094\>53261\>32711\>49601\>65840$         \\
 $3^3D$ & 22000  & $-2.05563\>63094\>53261\>32711\>49601\>65851$         \\
 & \cite{Drake92}& $-2.05563\>63094\>53261(4)$  & \\[1mm]
\hline\hline
\end{tabular}
\end{center}
\end{table}

\section{Inverse iteration method}

\begin{table}
\caption{Comparison of nonrelativistic energies of the ground state of a helium atom.}\label{compare1}
\begin{tabular}{l@{\hspace{5mm}}c@{\hspace{5mm}}r@{\hspace{8mm}}l}
\hline\hline
Author (year) & Ref. & $N$~~ & Energy (in a.u.) \\
\hline
Drake \emph{et al.} (2002)  & \cite{Drake02}     &  2358 & $-2.90372\>43770\>34119\>598311$ \\
Korobov (2002)              & \cite{Korobov02}   &  5200 & $-2.90372\>43770\>34119\>59831\>1159$ \\
Schwartz (2006)             & \cite{Sch06}       & 24099 &
                                           $-2.90372\>43770\>34119\>59831\>11592\>45194\>40444\>66969\>25310$ \\
Nakashima, Nakatsuji (2007) & \cite{Nakashima07} & 22709 & $-2.90372\>43770\>34119\>59831\>11592\>45194\>40444\>66969$ \\
this work                   &                    & 22000 & $-2.90372\>43770\>34119\>59831\>11592\>45194\>40443$ \\
\hline\hline
\end{tabular}
\end{table}

It was shown in Section 1 that the stationary Schr\"odinger equation is reduced to the generalized symmetrical  eigenvalue problem with the help of the Ritz procedure:
\begin{equation}\label{gse}
    Ax=\lambda Bx
\end{equation}
where $A$ is a symmetric matrix and $B$ is a symmetric positive definite matrix. The standard diagonalization procedure may be used to solve Eq. (\ref{gse}). In order to do that, matrix $B=L\cdot L^T$ is expanded into a product of upper and lower triangular matrices, and the problem is reduced to the standard symmetrical eigenvalue problem:
\begin{equation}
   A^{'}y=\lambda y,
\end{equation}
where
\begin{equation}
   A^{'}=L^{-1}AL^{-T},~~~~~ y=L^Tx.
\end{equation}

However, this method is too laborious ($\sim\!20\,N^3$ multiplication operations) and is less resistant to calculation errors. If  only  a  single  eigenvalue  (eigenvector)  is needed,   the   solution   may   be   obtained   efficiently ($\sim\!N^3/6$ multiplication operations) with the help of the inverse iteration method:
\begin{equation}
    (A-\mu)x_{k}^{(n+1)}=s^{(n)}x_k^{(n)},
\end{equation}
where  scalar  factor $s^{(n)}$  is  chosen  in  such  a  way  that $||{x_k^{(n+1)}}||=1$. If $\mu$ is close to exact eigenvalue $\lambda_k$ , vector sequence  $x_k^{(n)}$ converges  rapidly  to  exact  eigenvector $x_k$, and $\lambda_k^{(n)}=(x_k^{(n)}$, $Ax_k^{(n)})$ converges rapidly to exact value $\lambda_k$.
In order to illustrate this, one may assume, without a loss of generality, that matrix $A$ is a diagonal one. The solution  may  then  be  written  down  in  the  explicit form:
\begin{equation}\label{inviter}
    x_k^{(n)}=c_n\biggl(\biggl(\frac {\lambda_k-\mu}{\lambda_1-\mu}\biggl)^n u_1,..., u_k,...,\biggl(\frac{\lambda_k-\mu}{\lambda_n-\mu}\biggl)^n u_n\biggl)^T.
\end{equation}
It can be seen from Eqs.~(\ref{inviter}) that all components of vector $x_k^{(n)}$ (except  for $u_k$,  which  remains  equal  to unity) tend to zero under the given normalization conditions.  Practical  calculations  demonstrate  that  this method  is  also  the  most  resistant  to  rounding  errors (calculation errors).

\section{Results and discussion}

In Table \ref{converg1} we check the convergence of energy for the ground state of helium versus increasing basis sets of the variational expansion. The structure of "layers" of basis functions is very similar to what was used in our previous calculations \cite{Korobov02}, where it was explicitly published (see Table I in \cite{Korobov02}). In present case we optimized the variational basis with $N=10\,000$ functions and 8 layers. For the final calculation with $N=22\,000$ functions we used 12 layers, and for the last four layers the ends of intervals $[A_1,A_2]$ and $[B_1,B_2]$ grew exponentially: $A_1(j)=B_1(j)=10^{j-4}$, $A_2(j)=B_2(j)=10^{j-3}$ for $j=9,\dots,12$. Computations were performed in the duodecimal arithmetics (about 100 decimal digits). Programs of duodecimal precision were developed by our group in order to overcome the problem of the numerical instability of calculations at large values of $N$.

The final results of numerical calculations of the ionization energies for $S$, $P$ $D$, and $F$ states of a helium atom are presented in Table \ref{energies}. Variational parameters were optimized manually. It should be noted that the optimal variational parameters for different states differ significantly, and the calculation accuracy depends to a considerable extent (5-8 digits) on the particular choice of optimal variational parameters for a given bound state. Basis sets with $N=10\,000$ functions were used to optimize the variational parameters. When the non $S$ states listed in the table were calculated, 4 to 6 "layers" of basis functions were used, while for the $S$ states calculations were done in the similar way as for the ground state. The results in Table \ref{energies} are presented for two subsequent calculations with increasing basis sets, what allows to demonstrate convergent digits. The third line shows the results of variational calculations by Drake and Yan \cite{Drake92} performed in year 1992, where the Rydberg states (excluding $S$ states) of helium were studied. Comparison between two calculations demonstrates excellent agreement. The largest set for each particular state has been chosen by the reason that further increase of the basis gives rise to numerical instability of calculations within given duodecimal arithmetics. As may be seen numerical precision for triplet states is slightly higher, probably that is due to smaller effect of the logarithmic singularity. For higher orbital angular momentum states we have managed to achieve precision of 27-28 digits. Still that is the best known data for these states. All the calculations were performed on the Linux mainframe computers of our Laboratory.

For the ground state energy we compare our best obtained value with previously published results in Table \ref{compare1}. Indeed, explicit inclusion of the logarithmic singularity into a variational expansion may seriously improve precision of the results. On the other hand, with our variational basis function we can easily extend calculations to the states with excited electronic orbital as well as nonzero angular momentum states with large $L$.

The last two Tables \ref{conv2} and \ref{compare2} are devoted to the calculations of the single bound state in the negative hydrogen ion, H$^-$. In this case our numerical result for the energy is the most precise comparing to previous calculations of this quantity. In the work of Nakashima and Nakatsuji \cite{Nakashima07} the data presented in Table \ref{compare2} was claimed as convergent, presumably that indicates that the the free iterative complement interaction method possess some difficulties in the inner criterium to determine actual accuracy of the calculation.

Recently, we also studied applications of our method to the bound and resonant states in the hydrogen molecular ion H$_2^+$, which are supported by the ground electronic state $1s\sigma_g$ adiabatic potential \cite{MP18}. All the states up to $v=19$ vibration excited state and $L=41$ rotational state were accessible for very high precision calculations. That gives yet another evidence of great universality of the variational exponential expansion.

\begin{table}
\caption{Convergence of the nonrelativistic energy of the H$^-$ ion ground state. Exponential variational expansion.}\label{conv2}
\begin{tabular}{r@{\hspace{12mm}}l}
\hline\hline
 $N$~~ & ~~~~~~Energy (in a.u.) \\
\hline
 14000 & $-0.52775\,10165\,44377\,19659\,08145\,66747\,2$ \\
 18000 & $-0.52775\,10165\,44377\,19659\,08145\,66747\,55$ \\
 22000 & $-0.52775\,10165\,44377\,19659\,08145\,66747\,576$ \\
 26000 & $-0.52775\,10165\,44377\,19659\,08145\,66747\,5776$ \\
\hline\hline
\end{tabular}
\end{table}

\begin{table}
\caption{Comparison of nonrelativistic energies of the H$^-$ ion ground state.}\label{compare2}
\begin{tabular}{l@{\hspace{5mm}}c@{\hspace{5mm}}r@{\hspace{8mm}}l}
\hline\hline
Author (year) & Ref. & $N$~~ & ~~~~~~Energy (in a.u.) \\
\hline
Morgan \emph{et al.} (1990) & \cite{Morgan}      &  ---  & $-0.52775\>10165\>44375$ \\
Drake \emph{et al.} (2002)  & \cite{Drake02}     &  2276 & $-0.52775\>10165\>44377\>1965$ \\
Frolov (2006)               & \cite{Frolov06}    &  3700 & $-0.52775\>10165\>44377\>19659\>0$ \\
Nakashima, Nakatsuji (2007) & \cite{Nakashima07} &  9682 & $-0.52775\>10165\>44377\>19659\>08145\>66747\>511$ \\
this work                   &                    & 26000 & $-0.52775\>10165\>44377\>19659\>08145\>66747\>5776$ \\
\hline\hline
\end{tabular}
\end{table}

\section{Conclusions}

Variational wave functions of bound states are obtained by solving the Schrodinger equation for the quantum three-body problem with Coulomb interaction using a variational approach based on exponential expansion with the parameters of  exponents being chosen in a pseudorandom way. The results of these studies demonstrated that the energy values were accurate to 27--35 significant digits. In case of the negative hydrogen ion H$^-$ ground state we obtained the most accurate value as compared to the published data.

\section*{Acknowledgements} The work was supported by the Ministry of Education and Science Republic of Kazakhstan under grant IRN AP05132978, V.I.K. acknowledges support of the "RUDN University Program 5-100".

\end{document}